\pdfoutput=1
\documentclass[12pt,letterpaper,floatfix]{article}
\usepackage{amsmath,amssymb,bbm,color}
\usepackage{graphicx}
\usepackage{epsfig}
\usepackage{euscript}
\usepackage{pslatex}
\usepackage{xcolor}
\usepackage{fancybox}
\usepackage{enumerate}
\usepackage{widetext}
\usepackage{subfigure}
\definecolor{brightturquoise}{rgb}{0.03, 0.91, 0.87}
\definecolor{awesome}{rgb}{1.0, 0.13, 0.32}
\definecolor{armygreen}{rgb}{0.29, 0.33, 0.13}
\definecolor{aqua}{rgb}{0.0, 1.0,1.0}
\definecolor{maroon(html/css)}{rgb}{0.5, 0.0,0.0}
\definecolor{pinegreen}{rgb}{0.0, 0.47,0.44}
\definecolor{red-brown}{rgb}{0.65, 0.16,0.16}

\newcommand{\bhlumi}{{\tt BHLUMI}}
\newcommand{\bhwide}{{\tt BHWIDE}}
\newcommand{\babayaga}{{\tt BabaYaga}}
\newcommand{\order}[1]{${\cal O}(#1)$}

\usepackage[figuresright]{rotating}

\begin{document}

\begin{titlepage}
\begin{center}
\begin{flushleft}
{\small\bf BU-HEPP-19-01, Jan, 2019}
\end{flushleft}
\vspace{18mm}

{\bf\Large Path to the 0.01\% Theoretical Luminosity Precision Requirement for the FCC-ee (and ILC)}\\
\vspace{2mm}

{B.F.L. Ward$^a$\footnote{Speaker},~S. Jadach$^b$,~W. Placzek$^c$,~M. Skrzypek$^b$, and S.A. Yost,$^d$ }\\
{$^a$Baylor University, Waco, TX, USA}\\
{$^b$Institute of Nuclear Physics, Polish Academy of Sciences, Krakow, PL}\\
{$^c$Marian Smoluchowski Institute of Physics, Jagiellonian University,
Krakow, PL}\\
{$^d$The Citadel, Charleston, SC, USA}\\
\end{center}
\centerline{\bf Abstract}
We present pathways to the required theoretical precision for the luminosity targeted by the FCC-ee precision studies. We put the discussion in context by reviewing
briefly the situation at the time of LEP. We then present the current status and routes to the desired 0.01\% targeted by the FCC-ee (as well as by the ILC).
\vskip 25mm
\flushleft{ Talk presented at the International Workshop on Future Linear Colliders (LCWS2018),\\
Arlington, Texas, 22-26 October 2018. C18-10-22.}
\end{titlepage}
Our starting point will be the situation that existed at the end of LEP. At that time, the error budget for the \bhlumi4.04 MC~\cite{Jadach:1996is} used by all LEP collaborations to
simulate the luminosity process was calculated in Ref.~\cite{Ward:1998ht}. We reproduce this result here in Table ~\ref{tab:error99} for reference.
\begin{table}[!ht]
\centering
\begin{tabular}{|l|l|l|l|l|l|}
\hline 
    & \multicolumn{2}{|c|}{LEP1} 
              & \multicolumn{2}{|c|}{LEP2}
\\ \hline 
Type of correction/error
    & 1996
         & 1999
              & 1996
                   & 1999
\\  \hline 
(a) Missing photonic ${\cal O}(\alpha^2 )$~\cite{Jadach:1995hy,Jadach:1999pf} 
    & 0.10\%      
        & 0.027\%    
            & 0.20\%  
                & 0.04\%
\\ 
(b) Missing photonic ${\cal O}(\alpha^3 L_e^3)$~\cite{Jadach:1996ir} 
    & 0.015\%     
        & 0.015\%    
            & 0.03\%  
                & 0.03\% 
\\ 
(c) Vacuum polarization~\cite{Burkhardt:1995tt,Eidelman:1995ny} 
    & 0.04\%      
        & 0.04\%    
           & 0.10\%  
                & 0.10\% 
\\ 
(d) Light pairs~\cite{Jadach:1992nk,Jadach:1996ca} 
    & 0.03\%      
        & 0.03\%    
            & 0.05\%  
                & 0.05\% 
\\ 
(e) $Z$ and $s$-channel $\gamma$~\cite{Jadach:1995hv,Arbuzov:1996eq}
    & 0.015\%      
        & 0.015\%   
            &  0.0\%  
                & 0.0\% 
\\ \hline 
Total  
    & 0.11\%~\cite{Arbuzov:1996eq}
        & 0.061\%~\cite{Ward:1998ht}
            & 0.25\%~\cite{Arbuzov:1996eq}
                & 0.12\%~\cite{Ward:1998ht}
\\ \hline 
\end{tabular}
\caption{\sf
Summary of the total (physical+technical) theoretical uncertainty
for a typical calorimetric detector.
For LEP1, the above estimate is valid for a generic angular range
within   $1^{\circ}$--$3^{\circ}$ ($18$--$52$ mrads), and
for  LEP2 energies up to $176$~GeV and an
angular range within $3^{\circ}$--$6^{\circ}$.
Total uncertainty is taken in quadrature.
Technical precision included in (a).
}
\label{tab:error99}
\end{table}
In this table, we show the published works upon which the various error estimates are based as they are discussed in Ref.~\cite{Ward:1998ht}.\par
One way to address the 0.01\% precision tag needed for the luminosity theory error for the FCC-ee is to develop the corresponding improved version
of the \bhlumi. This problem is addressed recently in Ref.~\cite{Jadach:2018}, wherein the path to 0.01\% theory precision for the FCC-ee luminosity is 
presented in some detail. The results of this latter reference are shown in Table ~\ref{tab:lep2fcc}, wherein we also present the current state of the art for completeness, as it is discussed in more detail in Ref.~\cite{Jadach:2018}.
\begin{table}[ht!]
\centering
\begin{tabular}{|l|l|l|l|}
\hline
Type of correction~/~Error
        & Update 2018
                &  FCC-ee forecast
\\ \hline 
(a) Photonic $[{\cal O}(L_e\alpha^2 )]\; {\cal O}(L_e^2\alpha^3)$
        & 0.027\%
                &  $ 0.1 \times 10^{-4} $
\\ 
(b) Photonic $[{\cal O}(L_e^3\alpha^3)]\; {\cal O}(L_e^4\alpha^4)$
        & 0.015\%
                & $ 0.6 \times 10^{-5} $
\\
(c) Vacuum polariz.
        & 0.014\%~\cite{JegerlehnerCERN:2016}
                & $ 0.6 \times 10^{-4} $
\\
(d) Light pairs
        & 0.010\%~\cite{Montagna:1998vb,Montagna:1999eu}
                & $ 0.5 \times 10^{-4} $
\\
(e) $Z$ and $s$-channel $\gamma$ exchange
        & 0.090\%~\cite{Jadach:1995hv}
                & $ 0.1 \times 10^{-4} $
\\ 
(f) Up-down interference
    &0.009\%~\cite{Jadach:1990zf}
        & $ 0.1 \times 10^{-4} $
\\
(g) Technical Precision & (0.027)\% 
                & $ 0.1 \times 10^{-4} $
\\ \hline 
Total
        & 0.097\%
                & $ 1.0 \times 10^{-4} $
\\ \hline 
\end{tabular}
\caption{\sf
Anticipated total (physical+technical) theoretical uncertainty 
for a FCC-ee luminosity calorimetric detector with
the angular range being $64$--$86\,$mrad (narrow), near the $Z$ peak.
Description of photonic corrections in square brackets is related to 
the 2nd column.
The total error is summed in quadrature.
}
\label{tab:lep2fcc}
\end{table}

The key steps in arriving at Table~\ref{tab:lep2fcc} are as follows. The errors associated with the photonic corrections in lines (a) and (b) in the LEP results in Table~\ref{tab:error99} are due to effects which are known from Refs.~\cite{Jadach:1995hy,Jadach:1999pf,Jadach:1996ir} but which were not implemented into \bhlumi. In Table~\ref{tab:lep2fcc} we show what these errors will become after these known results are included in \bhlumi\ as discussed in Ref.~\cite{Jadach:2018}. Similarly, in line (c) of Table~\ref{tab:error99} the error is due to the uncertainty at the time of LEP on the hadronic contribution to the vacuum polarization for the photon at the respective momentum transfers for the luminosity process; in Table~\ref{tab:lep2fcc} we show the improvement of this error that is expected for the FCC-ee as discussed in Refs.~\cite{JegerlehnerCERN:2016,Jegerlehner:2017zsb,jegerlhnr-Fccee-2019}.
\par
Continuing in this way, in line (d) in Table~\ref{tab:lep2fcc} we show the expected~\cite{Jadach:2018} improvement, with reference to the LEP time for Table~\ref{tab:error99}, in the light pairs error for the FCC-ee . As we explain in Ref.~\cite{Jadach:2018}, the complete matrix element for the additional real $e^+e^−$ pair radiation should be
used, because non-photonic graphs can contribute as much as 0.01\% for the cut-off $z_{\rm cut} \sim 0.7$. This can be done with the MC generators developed for the $e^+e^-\rightarrow 4f$ processes for  LEP2 physics. With known methods~\cite{Jadach:2018}, the contributions of light quark pairs, muon pairs and non-leading, non-soft additional $e^+e^− +n\gamma$ corrections can be controlled such that the error on the pairs contribution is as given in line (d) for the FCC-ee. As noted, we also show the current state of the art~\cite{Jadach:2018} for this error in line(d) of Table~\ref{tab:lep2fcc}.
\par
Turning to line (e) in Table~\ref{tab:lep2fcc},  we show the improvement of the error on the Z and s-channel $\gamma$ exchange for the FCC-ee as well as its current state of the art.
In Ref.~\cite{Jadach:2018}, a detailed discussion is given of all of the six interference and three additional squared modulus terms  that result from the s-channel $\gamma$,  s-channel Z, and t-channel Z exchange contributions to the amplitude for the luminosity process. It is shown that, if the predictions of \bhlumi\ for the luminosity
measurement at FCC-ee are combined with the ones from \bhwide ~\cite{Jadach:1995nk} for
this Z and s-channel $\gamma$ exchange contribution, then the error in the second column of  line (e) of
Table~\ref{tab:lep2fcc} could be reduced to $0.01\%$. In order to reduce the uncertainty of this contribution
practically to zero we would include these $Z$ and $\gamma_s$ exchanges
within the CEEX~\cite{Jadach:2000ir} type matrix element at \order{\alpha^1} in \bhlumi. Here, CEEX stands for coherent exclusive exponentiation which acts at the level of the amplitudes
as compared the original Yennie--Frautschi--Suura~\cite{Yennie:1961ad}(YFS)  exclusive exponentiation (EEX) that is used in \bhlumi4.04 and that acts at the level of the squared amplitudes.
It is expected to be enough to add the EW corrections
to the LABH process in the form of effective couplings in the Born amplitudes. This leads to the error estimate shown in Table~\ref{tab:lep2fcc} in line(e) for the FCC-ee.
\par
In line(f) in Table~\ref{tab:lep2fcc} we show the estimate of the error on the up-down interference between radiation from the $e^-$ and $e^+$  lines. Unlike in LEP1, where it was negligible, for the FCC-ee this effect, calculated in Ref.~\cite{Jadach:1990zf} at ${\cal O}(\alpha^1)$, is 10 times larger and has to be included in the upgraded \bhlumi. Once this is done, the error estimate shown in line(f) for the FCC-ee obtains~\cite{Jadach:2018}.
\par
This brings us to the issue of the technical precision. In an ideal situation,
in order to get the upgraded \bhlumi\'s technical precision
at the level $10^{-5}$ for the total cross section and $10^{-4}$
for single differential distributions, one would need to compare
it with another MC program developed independently,
which properly implements the soft-photon resummation,
LO corrections up to \order{\alpha^3 L_e^3}, and
the second-order corrections with the complete \order{\alpha^2 L_e}.
In principle, an extension of a program like \babayaga ~\cite{CarloniCalame:2000pz,CarloniCalame:2001ny,Balossini:2006wc}, which is currently exact at NLO with a matched QED shower,
to the level of NNLO for the hard process, while
keeping the correct soft-photon resummation,
would provide the best comparison to the upgraded \bhlumi\ to establish the
technical precision of both programs at the $10^{-5}$
precision level%
\footnote{ The upgrade of the \bhlumi\ distributions will be
  relatively straightforward because its multi-photon 
  phase space is exact~\cite{Jadach:1999vf} for any number of photons.}.
During the intervening time period, a very good test of the technical precision of the upgraded \bhlumi\ would follow from the comparison between its results with EEX and CEEX matrix elements; for,
the basic multi-photon phase space integration module of \bhlumi\
was already well tested in Ref.~\cite{Jadach:1996bx}
and such a test can be repeated at an even higher-precision level.
\par
In summary,  we conclude that, with the appropriate resources, the path to $0.01\%$ precision for the FCC-ee luminosity (and the ILC luminosity) at the Z peak is open
via an upgraded version of \bhlumi.

\providecommand{\href}[2]{#2}




\begingroup\begin{thebibliography}{10}
\bibitem{Jadach:1996is}
S.~Jadach, W.~Placzek, E.~Richter-Was, B.~F.~L. Ward, and Z.~Was, ``{Upgrade of the Monte Carlo program \bhlumi\ for Bhabha scattering at low angles to version 4.04}'', {\em Comput. Phys. Commun.} {\bf 102} (1997) 229--251.
\bibitem{Ward:1998ht}
S.~Jadach, M.~Melles, B.~F.~L. Ward, and S.~A. Yost, ``{New results on the
  theoretical precision of the LEP / SLC luminosity}'', {\em Phys. Lett.} {\bf
  B450} (1999) 262--266,
\href{http://www.arXiv.org/abs/hep-ph/9811245}{hep-ph/9811245}
\bibitem{Jadach:1995hy}
S.~Jadach, M.~Melles, B.~F.~L. Ward, and S.~A. Yost, ``{Exact results on O
  (alpha) corrections to the single hard bremsstrahlung process in low angle
  Bhabha scattering in the SLC / LEP energy regime}'', {\em Phys. Lett.} {\bf
  B377} (1996) 168--176,
\href{http://www.arXiv.org/abs/hep-ph/9603248}{hep-ph/9603248}.
\bibitem{Jadach:1999pf}
S.~Jadach, M.~Melles, B.~F.~L. Ward, and S.~A. Yost, ``{New results on the
  precision of the LEP luminosity}'', {\em Acta Phys. Polon.} {\bf B30} (1999)
1745--1750.
\bibitem{Jadach:1996ir}
S.~Jadach and B.~F.~L. Ward, ``{Missing third order leading log corrections in
  the small angle Bhabha calculation}'', {\em Phys. Lett.} {\bf B389} (1996)
129--136.

\bibitem{Burkhardt:1995tt}
H.~Burkhardt and B.~Pietrzyk, ``{Update of the hadronic contribution to the QED
  vacuum polarization}'', {\em Phys. Lett.} {\bf B356} (1995)
398--403.

\bibitem{Eidelman:1995ny}
S.~Eidelman and F.~Jegerlehner, ``{Hadronic contributions to g-2 of the leptons
  and to the effective fine structure constant alpha (M(z)**2)}'', {\em Z.
  Phys.} {\bf C67} (1995) 585--602,
\href{http://www.arXiv.org/abs/hep-ph/9502298}{hep-ph/9502298}.
\bibitem{Jadach:1992nk}
S.~Jadach, M.~Skrzypek, and B.~F.~L. Ward, ``{Analytical results for low angle
  Bhabha scattering with pair production}'', {\em Phys. Rev.} {\bf D47} (1993)
3733--3741.

\bibitem{Jadach:1996ca}
S.~Jadach, M.~Skrzypek, and B.~F.~L. Ward, ``{Soft pairs corrections to low
  angle Bhabha scattering: YFS Monte Carlo approach}'', {\em Phys. Rev.} {\bf
  D55} (1997)
1206--1215.

\bibitem{Jadach:1995hv}
S.~Jadach, W.~Placzek, and B.~F.~L. Ward, ``{Precision calculation of the gamma
  - Z interference effect in the SLC / LEP luminosity process}'', {\em Phys.
  Lett.} {\bf B353} (1995)
349--361.

\bibitem{Arbuzov:1996eq}
A.~Arbuzov {\em et al.}, ``{The Present theoretical error on the Bhabha
  scattering cross-section in the luminometry region at LEP}'', {\em Phys.
  Lett.} {\bf B383} (1996) 238--242,
\href{http://www.arXiv.org/abs/hep-ph/9605239}{hep-ph/9605239}.
\bibitem{Jadach:2018}
S.~Jadach, W.~Placzek, M.~Skrzypek, B.~F.~L. Ward, and S.~A. Yost, {\em Phys. Lett.} {\bf B790} (2019) 314,
\href{http://www.arXiv.org/abs/1812.01004}{1812.01004}.
\bibitem{JegerlehnerCERN:2016}
F.~Jegerlehner, ``{$\alpha_{QED}(MZ)$ and future prospects with low energy e+e−
collider data }'',
\href{https://indico.cern.ch/event/469561/contributions/1977974/attachments/1221704/1786449/SMalphaFCCee16.pdf}{FCC-ee
Mini-Workshop, ”Physics Behind Precision” {\em
https://indico.cern.ch/event/469561/}}.
\bibitem{Montagna:1998vb}
G.~Montagna, M.~Moretti, O.~Nicrosini, A.~Pallavicini, and F.~Piccinini,
  ``{Light pair correction to Bhabha scattering at small angle}'', {\em Nucl.
  Phys.} {\bf B547} (1999) 39--59,
\href{http://www.arXiv.org/abs/hep-ph/9811436}{hep-ph/9811436}.
\bibitem{Montagna:1999eu}
G.~Montagna, M.~Moretti, O.~Nicrosini, A.~Pallavicini, and F.~Piccinini,
  ``{Light pair corrections to small angle Bhabha scattering in a realistic set
  up at LEP}'', {\em Phys. Lett.} {\bf B459} (1999) 649--652,
\href{http://www.arXiv.org/abs/hep-ph/9905235}{hep-ph/9905235}.
\bibitem{Jadach:1990zf}
S.~Jadach, E.~Richter-Was, B.~F.~L. Ward, and Z.~Was, ``{Analytical O(alpha)
  distributions for Bhabha scattering at low angles}'', {\em Phys. Lett.} {\bf
  B253} (1991)
469--477.
\bibitem{Jegerlehner:2017zsb}
F.~Jegerlehner, ``{Variations on Photon Vacuum Polarization}'',
\href{http://www.arXiv.org/abs/1711.06089}{1711.06089}.
\bibitem{jegerlhnr-Fccee-2019}
F.~Jegerlehner, talk in{\em 11th FCC-ee Workshop: Theory and Experiments}, Jan. 8-11, 2019, CERN, Geneva, Switzerland.
\bibitem{Jadach:1995nk}
S.~Jadach, W.~Placzek, and B.~F.~L. Ward, ``{\bhwide\ 1.00:${\cal O}(\alpha)$ YFS exponeniated Monte Carlo for Bhabha scattering at wide angles for LEP-1/SLC and LEP-2}'', {\em Phys. Lett.}{\bf B390} (1997) 298--308,
\href{http://www.arXiv.org/abs/hep-ph/9608412}{hep-ph/9608412}.
\bibitem{Jadach:2000ir}
S.~Jadach, B.~Ward, and Z.~W\c{a}s, ``{Coherent exclusive exponentiation for
  precision Monte Carlo calculations}'', {\em Phys. Rev.} {\bf D63} (2001)
  113009,
\href{http://www.arXiv.org/abs/hep-ph/0006359}{hep-ph/0006359}.
\bibitem{Yennie:1961ad} D.~R. Yennie, S.~C. Frautschi, and H.~Suura,''{The infrared divergence phenomena and high energy processes}'', {\em Annals Phys.}{\bf 13} (1961) 379--452.
\bibitem{CarloniCalame:2000pz}
C.~M. Carloni~Calame, C.~Lunardini, G.~Montagna, O.~Nicrosini, and
  F.~Piccinini, ``{Large angle Bhabha scattering and luminosity at flavor
  factories}'', {\em Nucl. Phys.} {\bf B584} (2000) 459--479,
\href{http://www.arXiv.org/abs/hep-ph/0003268}{hep-ph/0003268}.

\bibitem{CarloniCalame:2001ny}
C.~M. Carloni~Calame, ``{An Improved parton shower algorithm in QED}'', {\em
  Phys. Lett.} {\bf B520} (2001) 16--24,
\href{http://www.arXiv.org/abs/hep-ph/0103117}{hep-ph/0103117}.

\bibitem{Balossini:2006wc}
G.~Balossini, C.~M. Carloni~Calame, G.~Montagna, O.~Nicrosini, and
  F.~Piccinini, ``{Matching perturbative and parton shower corrections to
  Bhabha process at flavour factories}'', {\em Nucl. Phys.} {\bf B758} (2006)
  227--253,
\href{http://www.arXiv.org/abs/hep-ph/0607181}{hep-ph/0607181}.
\bibitem{Jadach:1999vf}
S.~Jadach, B.~F.~L. Ward, and Z.~Was, ``The precision {M}onte {C}arlo event
  generator {KK} for two- fermion final states in e+ e- collisions'', {\em
  Comput. Phys. Commun.} {\bf 130} (2000) 260--325, Program source available
  from http://jadach.web.cern.ch/,
\href{http://arXiv.org/abs/hep-ph/9912214}{hep-ph/9912214}.
\bibitem{Jadach:1996bx}
S.~Jadach and B.~F.~L. Ward, ``{Semianalytical third order calculations of the
  small angle Bhabha cross-sections}'', {\em Acta Phys. Polon.} {\bf B28}
  (1997)
1907--1979.
\end{thebibliography}\endgroup







\end{document}